\documentclass[pre,onecolumn,nofootinbib,superscriptaddress,showkeys,11pt]{revtex4}
\usepackage{comment}
\usepackage{graphicx}
\usepackage{graphicx,subfigure}
\usepackage{amsmath,amssymb,amsfonts}
\DeclareMathOperator{\sech}{sech}
\usepackage[usenames]{color}
\usepackage{bm}
\usepackage{epsfig}
\usepackage{dcolumn}
\usepackage{xspace}

\begin{document}
\title{Instabilities, thermal fluctuations, defects and dislocations in the crystal-$R_I$-$R_{II}$ rotator phase transitions of n-alkanes}
\author{Soumya Kanti Ganguly}
\affiliation{Ongil Private Limited,Chennai 600113, India}
\author{Prabir K. Mukherjee\protect\footnote
{E-mail address: pkmuk1966@gmail.com}}
\affiliation{Department of Physics, Government College of Engineering and Textile Technology, 12 William Carey Road, Serampore, Hooghly-712201, India}


\begin{abstract}
The theoretical study of instabilities, thermal fluctuations, and topological defects in the crystal-rotator-I-rotator-II ($X-R_{I}-R_{II}$) 
phase transitions of n-alkanes has been conducted. First, we examine the nature of the $R_{I}-R_{II}$ phase transition in nanoconfined alkanes. 
We propose that under confined conditions, the presence of quenched random orientational disorder makes the $R_{I}$ phase unstable. This 
disorder-mediated transition falls within the Imry-Ma universality class. Next, we discuss the role of thermal fluctuations in certain rotator 
phases, as well as the influence of dislocations on the $X-R_I$ phase transition. Our findings indicate that the Herringbone order in the 
$X$-phase and the Hexatic order in the $R_{II}$-phase exhibit quasi-long-range characteristics. Furthermore, we find that in two dimensions,
the unbinding of dislocations does not result in a disordered liquid state.

\keywords{Rotator phase, confinement, instabilities, dislocations, topological defects
phase transition}
\end{abstract}


\maketitle

\section{Introduction}\label{Intro}

The crystalline form of n-alkanes exhibits a layered structure made up of bilayer stacks of lamellas. 
When these n-alkanes are heated, they reach a point where the molecules are well-ordered in terms of 
their translations but exhibit rotational disorder. This newly formed phase is referred to as the 
rotator phase. To date, five distinct rotator phases have been identified in n-alkanes. Due to the 
unique and unusual properties of these rotator phases, they have become a significant area of research 
over the past few decades \cite{sirota1,muk1}.

\begin{figure}[ht]
\includegraphics[width=0.5\textwidth]{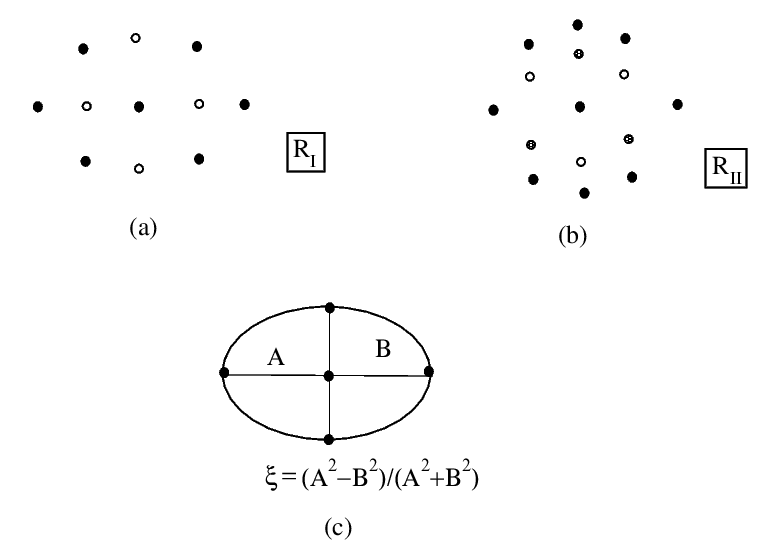}
\caption{Schematic showing the symmetries of the (a) $R_{I}$ and (b) $R_{II}$ phases.
The chain end positions in the second layer of the $R_I$ phase is represented by unfilled 
circles. Third layer of the $R_{II}$ phase is represented by gray circles. 
(c) The definition of lattice distortion parameter ($\xi$).}
\label{fig1}
\end{figure}

The low-temperature crystalline phase ($X$) of n-alkanes has a layered structure characterized by the 
bilayer stacking of lamellas \cite{Do1,Do2,Do3,ungar1,ungar2}. Additionally, the $X$ phase exhibits 
either orthorhombic (distorted hexagonal) or triclinic packing within each layer, along with a long-range 
herringbone arrangement of the rotational degrees of freedom of the molecular backbones \cite{sirota1}. 
Among the five different rotator phases, the rotator-I ($R_I$) and rotator-II ($R_{II}$) phases are the 
most extensively studied. The $R_I$ phase features bilayer stacking in a rectangularly distorted hexagonal 
lattice, with the molecules remaining untilted relative to the layers. In contrast, the $R_{II}$ phase is 
characterized by trilayer stacking within a rectangular hexagonal lattice, with the molecules also untilted 
with respect to the layers.

The $X-R_I$ phase transition is classified as first-order due to the significant hysteresis observed during 
supercooling \cite{Sirota:93, Sirota:94}. However, x-ray diffraction studies indicate that the $X-R_I$ phase 
transition is weakly first-order. Research has shown that a substantial number of chain-end gauche bond defects 
play a crucial role in the rotator phase transitions of n-alkanes \cite{maroncelli,jarrett, goworek, muk3}. 
Sirota et al. \cite{Sirota:93,Sirota:94} conducted extensive studies on the $R_I-R_{II}$ phase transition through 
x-ray scattering and Differential Scanning Calorimetry (DSC). The findings reveal a jump in the lattice distortion 
order parameter, a sharp peak in the heat capacity curve, and significant hysteresis during supercooling, all of 
which indicate the first-order nature of the $R_{I}-R_{II}$ phase transition \cite{Sirota:93, Sirota:94}. The complex 
interactions between intramolecular and intermolecular instabilities contribute to the displacive structure observed 
in the $R_I-R_{II}$ phase transition. Moreover, confinement has dramatic effects on the rotator phase transitions in 
n-alkanes.

Studies on rotator phase transitions under confinement conditions provide valuable insights into various thermophysical 
and interfacial phenomena \cite{jiang08,jiang09,jiang10,fu1,fu2,kumar,zammit1}. Huber et al. \cite{huber1,huber2} 
experimentally demonstrated that confinement stabilizes the rotator phase of alkanes, even when it is not present in 
the bulk state. Henschel et al. \cite{henschel} observed that solid alkanes can be thermodynamically stable in porous 
silicon, but become unstable if the pore walls are slightly oxidized. The $R_{I}-R_{II}$ phase transition is known to 
be first-order in nature, with confinement effects that are relatively weak \cite{zammit1}. Mukherjee \cite{muk2} 
discussed the influence of quenched disorders during the $R_{I}-rotator-V (R_{V})$ phase transition and how it mediates 
a second-order phase transition. It was assumed that the disorder function exhibited Gaussian correlated behavior in space, 
characterized by a correlation length $\Lambda$. The $R_{II}$-Liquid phase transition was experimentally investigated by 
Zammit et al. \cite{zammit1,zammit2} in pure and binary mixtures of alkanes. They observed a single peak in both the 
specific heat and latent heat of the pure material, indicating the first-order nature of the $R_{II}$-Liquid phase transition.

In this research, we initially examine the impact of quenched disorder during the transition from the $R_{I}$ phase to the 
$R_{II}$ phase. We propose that in a restricted scenario, the existence of quenched random orientational disorder in low 
dimensions will make the $R_{I}$ phase unstable. This kind of phase transition, influenced by random disorder, falls under 
the Imry-Ma universality class \cite{Imry}. Then, we discuss the significance of thermal fluctuations and dislocations in 
the different rotator phases. Our findings suggest that the release of dislocations plays a partial role in the transition 
from the $X$ phase to the $R_{I}$ phase in n-alkanes. We argue that the abrupt nature of the $X-R_{I}$ phase transition may 
be attributed to the dislocations occurring in all three dimensions, spanning across the bi-layers.


\section{$R_{I}-R_{II}$ phase transition in nanoconfinement}\label{Sec1}

\begin{figure}[ht]
\centering
\subfigure[]{\includegraphics[width=0.42\textwidth]{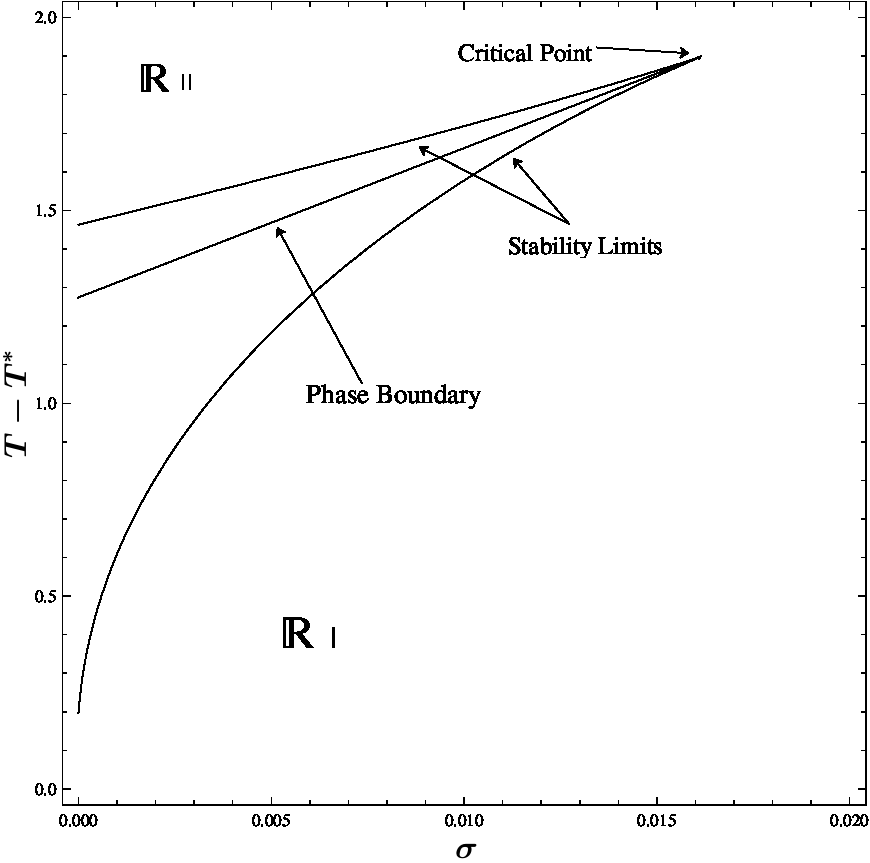}}
\subfigure[]{\includegraphics[width=0.42\textwidth]{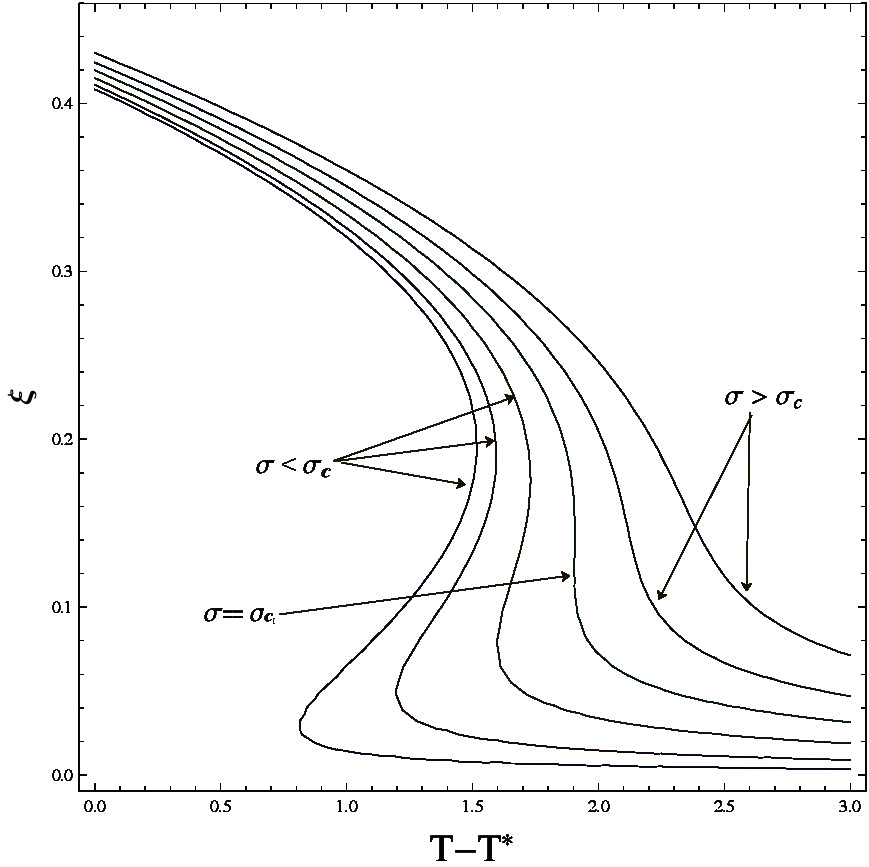}}
\caption{Phase diagrams of the $R_{I}-R_{II}$ phase transition in nanoconfinement: (a) The random orientational 
field ($\sigma$) vs. temperature $T-T^{*}$(K); (b) The distortion order parameter ($\sigma$) vs. temperature 
$T-T^{*}$ (K) for different $\sigma$ in the $R_{I}$ phase. The parameters $\alpha_{0} = 0.2$ J/cm$^3K$, 
$\beta = 2.75$ J/cm$^3$ units, and $\gamma = 6.88$ J/cm$^3$.} 
\label{fig2-3}
\end{figure}

Alkanes can be confined in materials such as Vycor glass, microcapsules, and Polytetrafluoroethylene (PTFE) matrixes. 
When alkanes are confined, their behavior is influenced by the competition between correlation lengths and the finite 
size of the system. This confinement reduces the dimensionality of the system and alters the surface forces at play. 
Research has shown that, in such constrained environments, the static or dynamic correlation lengths of alkanes are 
limited by the size of the confining structure. This theoretical limitation can be analyzed by applying a unidirectional 
field that is imposed by the porous solid.

The lattice distortion parameter, denoted as $\xi$, serves as an order parameter for describing the $R_I-R_{II}$ phase transition 
\cite{sirota1,muk1,Kaganer:99}. It is defined as \cite{sirota1,Kaganer:99} $\xi = (A^{2} - B^{2})/(A^{2} + B^{2})$, where $A$ and $B$ 
represent the lengths of the major and minor axes of an ellipse that passes through the six nearest neighbors. The value of $\xi$ 
is non-zero in the $R_I$ phase, indicating a distortion, while it is zero in the $R_{II}$ phase, indicating no distortion. The 
Ginzburg-Landau energy functional in $d$ dimensions for the $R_I-R_{II}$ phase transition is given by

\begin{equation}\label{Confinement1A}
H[\xi,\sigma] = \int d^{d}\boldsymbol{x} \bigg[ \frac{c}{2} \vert \vec{\nabla}\xi(\boldsymbol{x})\vert^{2} + 
\frac{\alpha}{2}\xi^{2}(\boldsymbol{x}) - \frac{\beta}{3}\xi^{3}(\boldsymbol{x}) + \frac{\gamma}{4}\xi^{4}(\boldsymbol{x}) 
- \sigma(\boldsymbol{x})\xi(\boldsymbol{x})\bigg]
\end{equation}

The last term represents the orientational field $\sigma$ that is coupled with $\xi$. The field $\sigma$ can be viewed as a 
unidirectional field imposed by porous silicon, which depends on the alkane and the confinement solid (pinning potential). 
The coefficient $\alpha = \alpha_{0}(T - T^{*})$, where $T^{*}$ is the absolute stability limit temperature of the $R_{II}$ phase. 
The coefficients $\alpha_{0},\beta,\gamma$ and $c$ are positive. The equilibrium phase diagram for the $R_{I}-R_{II}$ transition 
under nanoconfinement can be derived using the mean-field approximation. To achieve this, we will disregard fluctuations ($c = 0$) 
and assume a uniform value of the order parameter $\xi(\boldsymbol{x}) = \xi$, throughout the system. At equilibrium, the free energy 
will have a minimum if 

\begin{equation}
\frac{\partial H[\xi,\sigma]}{\partial \xi} = \alpha \xi - \beta \xi^{2} + \gamma \xi^{3} - \sigma = 0
\label{cond1}
\end{equation}

We numerically solve the above equation to calculate the thermodynamic phase boundary and stability limits of the $R_{I}$ and $R_{II}$ 
phases under confinement. We find that the $R_{I}-R_{II}$ phase transition terminates at a critical point within the confined environment 
(FIG.\ref{fig2-3}(a)). As the parameter $\sigma$ increases, we observe that the first-order $R_{I}-R_{II}$ phase transition transforms into 
a second-order transition FIG.\ref{fig2-3}(b). Therefore, under confined conditions, the $R_{I}-R_{II}$ phase transition can be second-order. 


\section{Confined $R_{I}-R_{II}$ transition in low dimensions, domain walls, and the Imry-Ma effect}\label{Sec2}

\begin{figure}[!ht]
\centering
\includegraphics[width=0.45\textwidth]{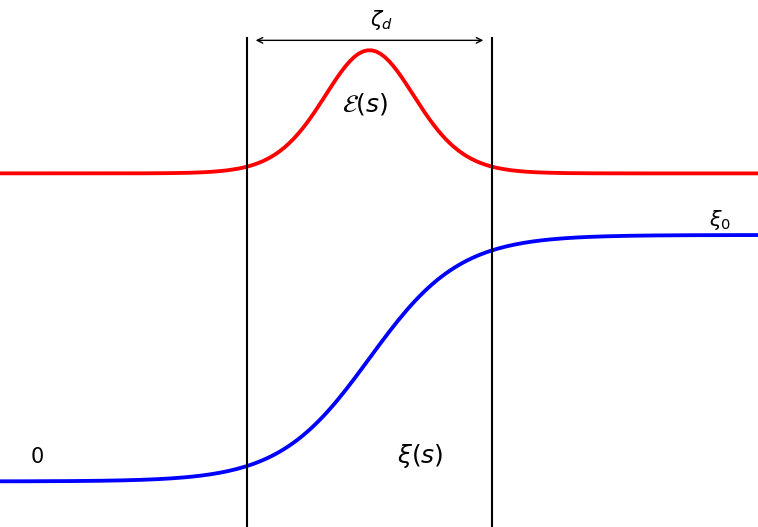}
\caption{Plots for the energy density (red) and the order-parameter (blue) for the domain wall 
(See equations \eqref{AppendixA1g} and \eqref{AppendixA1f} in \ref{AppB}).}
\label{Domain_Wall1}
\end{figure}

In the context of magnetic systems, the effect of static random external fields was
first studied by Imry and Ma \cite{Imry}. They chose the $N$-dimensional vector spin
($O(N)$) model as their model system in $d$ dimensions. The ground state and the thermal 
properties of many physical systems can be conveniently modelled using the $O(N)$ model. 
The dimension of the order parameter and the physical dimensions of the system are crucial 
factors in the Imry-Ma theory. The $N = 1$ case corresponds to the well-known Ising and the 
binary liquid model \cite{Stanley}, where the topological defects manifest as simple domain 
walls \cite{Mermin, Volovik} (FIG.\ref{Domain_Wall1}, blue curve). Considering the $N = 1$ 
case as our model system, we will see how the Imry-Ma theory can explain an unstable $R_{I}$ 
phase in the presence of static random confining fields. The $N = 2$ case will be examined in 
\ref{Sec4} and \ref{AppC}.

Based on the local values of the lattice distortion parameter $\xi$, the $R_{I}$ and $R_{II}$ phases 
can be compared to the two phases of a binary liquid system, which are separated by domain walls. 
This differs from the approach taken by Wentzel, who defined local Ising and Potts order parameters 
to characterize the orthorhombic and monoclinic crystalline phases of rotator systems \cite{Wentzel}. 
The total energy of the system is the sum of the surface energy, and the bulk energy.  

Let the value of the order parameter in the $R_{I}$ phase be $\xi_{0}$. The orientational field 
$\sigma(\boldsymbol{x})$, which acts as static quenched disorder, has the following statistical properties:

\begin{eqnarray}\label{Confinement1B}
&&\langle \sigma(\boldsymbol{x}) \rangle = 0, \quad \langle \sigma(\boldsymbol{x}) 
\sigma(\boldsymbol{x^{\prime}}) \rangle = f(\boldsymbol{x}-\boldsymbol{x^{\prime}}), \\ \nonumber
&&\langle \sigma(\boldsymbol{k}) \sigma(\boldsymbol{k^{\prime})} \rangle = \tilde{f}(\boldsymbol{k})\delta(\boldsymbol{k}-\boldsymbol{k^{\prime}}) 
\quad (\textrm{reciprocal space}).
\end{eqnarray}

Where the spatial correlation properties of $\sigma$ is given by f($\boldsymbol{x})$ whose particular form 
is not important for the present purpose. Let us denote the fluctuations by $\langle \sigma^{2} \rangle$. 
If we represent $L_{\Delta}$ as the typical size of a domain wall and $V$ as the volume of the system in $d$ 
dimensions, then there are typically about ${V/L_{\Delta}^{d}}$ domains present within the system. If the 
system has a critical dimension $d_{c}$, then for $d < d_{c}$, when the system in its lowest energy state, 
we will have approximately 

\begin{equation}
\frac{L_{\Delta}}{a} = \bigg(\frac{c \xi_{0}d_{c}}{2d\sqrt{\langle \sigma^{2} \rangle}}\bigg)^{\frac{2}{2-d_{c}}}
\end{equation}

number of defects (See \eqref{R12B}-\eqref{R12H} in in \ref{AppB}). Where $a$ is the lattice constant. The stability or the lack 
thereof of the $R_{I}$ phase will completely depend upon the outcome of the competition between the domain wall energy 
and the bulk energy. As mentioned, the $R_{I}-R_{II}$ transition is described by a scalar order parameter. For $d > d_{c}$, 
the surface energy is the dominant factor, the size of the domain wall becomes infinite, resulting in the system remaining 
in the $R_{I}$ state at any finite temperature. Conversely, if the random field is dominant, then the $R_{II}$ state becomes 
the lowest energy state, causing the size of the domain wall to shrink to the lattice spacing. In the present case $d_{c} = 2$, 
therefore, according to Imry and Ma, in $d = 2$, any amount of random confining field will destroy the $R_{I}$ phase.

It is important to note that the strong cubic non-linearity of the Landau-Ginzburg energy functional can lead to a first-order 
phase transition. During this transition, the system may become trapped in an ordered metastable state. The domain walls 
(FIG.\ref{Domain_Wall1}, blue curve) are fundamentally characterized by a length scale which is determined by the mean-field 
correlation length $\zeta_{d} = (c/\alpha)^{1/2}$ (see \eqref{AppendixA1f} in \ref{AppC}). For length scales larger than $\sqrt{2}\zeta_{d}$, 
the energy density (see \eqref{AppendixA1g} in \ref{AppC}) of these domain walls decreases rapidly (FIG.\ref{Domain_Wall1}, red curve). 
For practical purposes, $\zeta_{d}$ serves as the natural length scale for experimental observations. Furthermore, the surface tension of 
the domain wall $\sigma_{d}$ is related to the correlation length, and has the following expression.

\begin{equation}\label{AppendixA1h}
\sigma_{d} = \frac{2\sqrt{2}}{3}\alpha\zeta_{d}\xi^{2}_{0}. 
\end{equation}

\section{Role of thermal fluctuations and correlations in the $X$ and the $R_{II}$ phase}\label{Sec3}

\begin{figure}[!ht]
\centering
\includegraphics[width=0.32\textwidth]{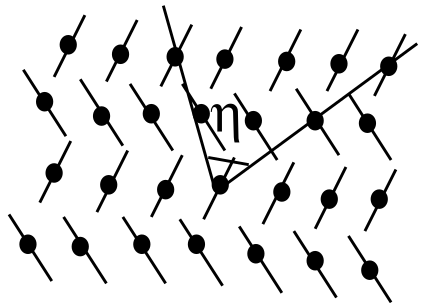}
\caption{The phase angle $\eta$ of the Herringbone order parameter.}
\label{HerringboneX}
\end{figure}

It was observed by Muller \cite{muller} that certain paraffins undergo a transition from a solid
crystalline state to a rotator phase just a few degrees below their melting temperature. In alkanes,
as the temperature increases, different rotator phases appear successively as a result of the gradual
breakdown of orientational order. The transition between the different rotator phases correspond to the
ireversible breakdown of certain symmetries. In this section we will examine the role of thermal
fluctuations and correlations in the different ordered states of the $X$-phase and the $R_{II}$-phase
of the crystalline n-alkane system.

The $X$-phase in n-alkane is characterized by the co-existence of the crystalline order, the lattice
distortion state $\xi$, and the herringbone order \cite{Do1,sirota1}. In the $R_{I}$ phase, the same
lattice distortion $\xi$ is maintained with the simultaneous disappearence of the herringbone order.
According to the vibrational theory of crystalline solids, the Goldstone modes which are the low
energy or long wavelength excitations are known as phonons. They are associated with the broken
translational symmetry in solids. The modes of vibration in the $i$-th direction is denoted by the
displacement fields $u_{i}$. For isotropic solids there are two elastic constants $\mu$ and $\lambda$
(Lame' coefficients), and the thermal fluctuations of the displacement fields is an indirect measure of
crystalline order in the system. In two dimensions, these fluctuations are given by
(See \eqref{AppendixA1}-\eqref{AppendixA6} in \ref{AppD} and \ref{AppE})

\begin{equation}\label{X-phase5}
\langle [u_{i}(\boldsymbol{x}) - u_{i}(\boldsymbol{0})]^{2} \rangle
= \frac{3\mu+\lambda}{\mu(2\mu+\lambda)}\frac{\ln(\vert \boldsymbol{x}\vert/a)}{\pi}.
\end{equation}

If $\boldsymbol{x}_{0}$ be the lattice vector and $\boldsymbol{G}$ be any reciprocal lattice vector, then
$\boldsymbol{G.x_{0}} = 2n\pi$, where $n$ is an integer. The order parameter for the crystalline solids
is its density function in the reciprocal space $\rho_{\boldsymbol{G}}(\boldsymbol{x}) = \exp[i \boldsymbol{G}.\boldsymbol{u}(\boldsymbol{x})]$.
In two-dimensions or three-dimensions with stacked layers, the density correlation function obeys a power
law given below (See \eqref{AppendixA6}-\eqref{AppendixA7} in \ref{AppC}).

\begin{eqnarray}\label{X-phase6}
\langle \rho_{\boldsymbol{G}}(\boldsymbol{x}) \rho^{*}_{\boldsymbol{G}}(\boldsymbol{0})\rangle
&\approx& \bigg(\frac{a}{\vert \boldsymbol{x}\vert}\bigg)^{n_{\boldsymbol{G}}}. \\ \nonumber
\textrm{Where,} \quad
n_{\boldsymbol{G}} &=& \frac{\vert\boldsymbol{G}\vert^{2}(3\mu+\lambda)}{4\pi\mu(2\mu+\lambda)}
\end{eqnarray}

is the exponent. Since the Bragg peaks are the measurable quantities during diffraction experiments,
at finite $T$ and close to the reciprocal lattice vector $\boldsymbol{G}$, the structure factor is given by

\begin{eqnarray}
S(\boldsymbol{k} \approx \boldsymbol{G}) &\propto&
\int d^{2}\boldsymbol{x}e^{i\boldsymbol{(k-G).x}}\bigg(\frac{a}{\vert \boldsymbol{x}\vert}\bigg)^{n_{\boldsymbol{G}}} \\ \nonumber
&=& \frac{a}{\vert \boldsymbol{k}-\boldsymbol{G} \vert^{2-n_{\boldsymbol{G}}}}
\end{eqnarray}

As the temperature $T$ and the magnitude of $\boldsymbol{G}$ increase, the Bragg peaks exhibit a power law singularity. 
The low-temperature $X$ phase is characterized by Bragg reflection. Coexisting with this phase is the herringbone 
ordering, which represents a specific type of crystallization. The herringbone order is described by a complex scalar 
field order parameter, $\Phi(\boldsymbol{x}) = \phi e^{i2\eta(\boldsymbol{x})}$. Here $\eta$ is the tilt-angle with respect 
to the local Herringbone axis and a reference direction \cite{bruinsma,Kaganer:99} (FIG.\ref{HerringboneX}). Local distortions 
in the displacement fields $\boldsymbol{u}(\boldsymbol{x})$ can cause a rotation in the Herringbone phase angle $\eta(\boldsymbol{x})$.

\begin{equation}\label{Herringbone1}
\eta(\boldsymbol{x}) = -\frac{1}{2}\bigg(\frac{\partial u_{i}(\boldsymbol{x})}{\partial x_{j}} - \frac{\partial u_{j}(\boldsymbol{x})}{\partial x_{i}} \bigg) =
-\frac{1}{2} \boldsymbol{\hat{e}}_{k}.[\boldsymbol{\nabla} \times \boldsymbol{u}(\boldsymbol{x})]_{k}
\end{equation}

Using this we can calculate the fluctuations in $\Phi(\boldsymbol{x})$ by calculating the correlation function

\begin{equation}\label{Herringbone2}
\langle \Phi(\boldsymbol{x}) \Phi^{*}(\boldsymbol{0})\rangle \approx \phi^{2}\exp\bigg(-\frac{1}{a^{2}\mu}\bigg)
\end{equation}

We find that in the long distance limit the spatial correlations are independent of $\vert\boldsymbol{x}\vert$.
(\eqref{AppendixB1} in \ref{AppC}). However, the spatial correlations have a temperature
dependence, since $\mu \propto T^{-1}$.

Like, the Herringbone order in the $X$-phase, the $R_{II}$-phase is characterized by the hexatic order where
the order parameter is also a complex scalar field $\Psi(\boldsymbol{x}) = \psi e^{i6\theta(\boldsymbol{x})}$.
The thermal fluctuations in this case are given by

\begin{equation}\label{Hexatic1}
\langle \Psi(\boldsymbol{x}) \Psi^{*}(\boldsymbol{0})\rangle \approx \psi^{2}\exp\bigg(-\frac{9}{a^{2}\mu}\bigg),
\end{equation}

which also show a $T$ dependence in the long range limit. Thus the Herringbone order in the $X$-phase and the
Hexatic order in the $R_{II}$-phase may not have true long range order but quasi-long range order.

\section{Plausible Imry-Ma instability effects in the Herringbone and the $R_{II}$ phases}\label{Sec4}

\begin{figure}[ht]
\includegraphics[width=0.44\textwidth]{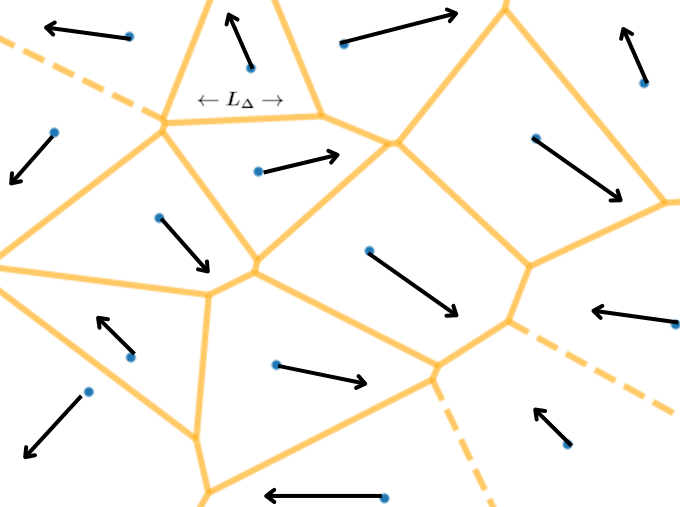}
\caption{Imry-Ma domain walls for Herringbone and $R_{II}$-like rotator models, where $L_{\Delta}$ is the typical size of the walls.}
\label{fig_O2}
\end{figure}

In sections \ref{Sec2} and \ref{Sec3}, we saw that the Herringbone and the $R_{II}$ phases can be characterized by the 
orientational order parameter which can take the general form $\Theta(\boldsymbol{x}) = \Theta_{0} e^{i\theta(\boldsymbol{x})}$. 
This system shares the same universal properties as the $O(2)$ (N = 2), or the XY-model. The topological defects in these systems 
are vortices or vortex-like excitations, leading to a Berezinskii-Kosterlitz-Thouless transition \cite{Berezinskii},\cite{KT}. 
If there is a confining field $h(\boldsymbol{x})$ similar to $\sigma(\boldsymbol{x})$ with the same set of statistical properties 
\eqref{Confinement1B}, then according to the Imry-Ma criterion, the critical dimension $d_{c} = 4$ (see \eqref{XYmodel5} in \ref{AppC}). 
Consequently, in $d = 3$, any amount of quenched disorder will lead the system to a disordered phase. However, we are currently 
unaware of the existence of such quenched disorders in three dimensions. Similar to the domain walls discussed in \ref{Sec2}, 
rotator systems will also exhibit their own domain walls, as shown in FIG.\ref{fig_O2}. If $\rho_{\Theta}$ represents the stiffness 
constant of the system and $\langle h^{2} \rangle$ indicates the variance of the external field, then for $d < d_{c}$ the lowest 
energy state of the system will have approximately

\begin{equation}\label{ImryMaXY1}
\frac{L_{\Delta}}{a} = \bigg(\frac{2\pi^{2}\rho_{\Theta}d_{c}}{\kappa(T)d\Theta_{0}\sqrt{\langle h^{2} \rangle}}\bigg)^{\frac{2}{2-d_{c}}}
\end{equation}

number of defects. Where $\kappa(T)$ is a temperature dependent parameter (See \eqref{XYmodel6} in \ref{AppC}). Furthermore, 
depending upon the strength of the confining field, the domain walls will have a characteristic length given by 

\begin{equation}\label{ImryMaXY2}
\lambda_{0} = \sqrt{\frac{\rho_{\Theta}}{\Theta_{0}h_{0}}}. 
\end{equation}

Where $h_{0}$ is the amplitude value of the random field (See \eqref{XYmodel7} in \ref{AppC}). 

\section{The role of crystal defects in the $X-R_{I}$ phase transition}\label{Sec5}

\begin{figure}[ht]
\centering
\subfigure[]{\includegraphics[width=0.2\textwidth]{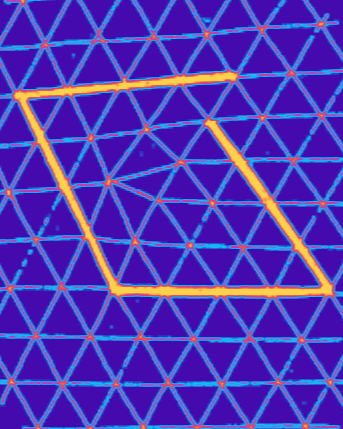}} \qquad
\subfigure[]{\includegraphics[width=0.4\textwidth]{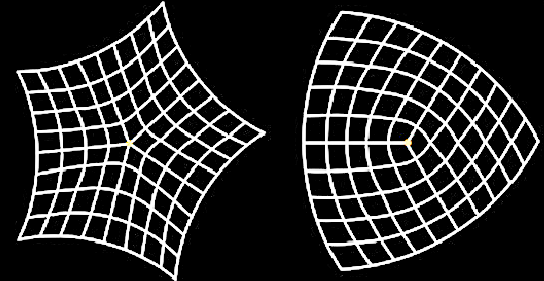}} 
\caption{Schematics of TDs in regular lattice systems: (a) Elementary dislocation in a triangular lattice. 
It occurs when the Burgers' vector, which indicates how much the path around a singularity fails to close, 
is present. In a perfect lattice, this path would form a closed loop; (b) Disclinations in a square lattice.}  
\label{Defects}
\end{figure}

In the previous section we looked at the effect of thermal fluctuations in the
different ordered phases of the crystalline n-alkane system. The order parameter 
fields which characterize broken continuous symmetry phases are represented by 
smooth continuous functions of $\boldsymbol{x}$. There can arise situations 
e.g. mechanical, thermal stress, when these fields may suffer from discontinuities 
which cannot be removed by any amount of smooth deformation. The presence of these 
defects affects the topology of the system and hence they are known as topological 
defects (TD). Such defects will have an inner core region where the order parameter 
vanishes and a far field region where the field changes slowly. Depending upon the 
nature of the broken symmetry, the TD's can have different names. For example, in 
crystalline solids where there are both broken translational and rotational symmetry, 
the TDs are known as dislocations and disclinations respectively (FIG.\ref{Defects}). 
Similarly, if the Herringbone order in $X$-phase, or the Hexatic order in the 
$R_{II}$-phase can be identified with the $XY$-model order parameter then the TDs in 
these phases will be vortices. In the present section we will focus on the role of 
dislocations in inducing the $X-R_{I}$ transition. For a dislocated crystal, the singular 
part of the displacement vector fields give rise to TDs. The displacement vector fields 
can be decomposed into a regular part $u_{i}^{(R)}$, and a singular part $u_{i}^{(S)}$.

\begin{equation}\label{Defect2C}
u_{i}(\boldsymbol{x}) = u_{i}^{(R)}(\boldsymbol{x}) + u_{i}^{(S)}(\boldsymbol{x}).
\end{equation}

Along any closed circuit $\Gamma$,

\begin{equation}\label{Defect2D}
\oint_{\Gamma} du_{i}^{(R)} = 0 \quad \textrm{and} \quad \oint_{\Gamma} du_{i}^{(S)} = b_{i}.
\end{equation}

Where $b_{i}$ is the Burgers' vector in the $i$-th direction. The energy for the dislocations in two
dimensions has the following expression \cite{Chaikin}

\begin{eqnarray}\label{Defect2E}
\frac{H_{D}}{T} &=&
-\tilde{\kappa}\sum_{\alpha<\alpha'} b_{i}(\boldsymbol{x}_{\alpha})U_{ij}(\boldsymbol{x}_{\alpha}-\boldsymbol{x}_{\alpha'})b_{j}(\boldsymbol{x}_{\alpha'})
+ \frac{E_{c}}{T}\sum_{i,\alpha} b^{2}_{i}(\boldsymbol{x}_{\alpha}). \\ \nonumber
\textrm{Where,} \quad
U_{ij}(\boldsymbol{x}) &=&
\frac{1}{2\pi}\bigg[\delta_{ij}\ln\bigg( \frac{\vert\boldsymbol{x}\vert}{a_{c}}\bigg) - \frac{x_{i}x_{j}}{\vert\boldsymbol{x}\vert^{2}} \bigg],
\quad
\tilde{\kappa}a^{2} = \frac{2\mu(\mu+\lambda)}{(2\mu+\lambda)},
\end{eqnarray}

and $E_{c}$ is the core energy. Additionally, we will have the defect neutrality condition $\sum_{\boldsymbol{x}} b_{i}(\boldsymbol{x}) = 0$. 
Similar to \eqref{Herringbone1} in the previous section, the singular part of the displacement fields can induce singular distortions in the 
Herringbone phase rotators. 

\begin{eqnarray}\label{X-RI-phase1}
\tilde{\eta}(\boldsymbol{x}) &=& -\frac{1}{2} \boldsymbol{\hat{e}}_{k}.[\boldsymbol{\nabla} \times \boldsymbol{u}^{(S)}(\boldsymbol{x})]_{k}
= \frac{1}{2\pi} \sum_{\alpha} 
\frac{\boldsymbol{b}(\boldsymbol{x}_{\alpha}).(\boldsymbol{x} - \boldsymbol{x}_{\alpha})}{\vert\boldsymbol{x}-\boldsymbol{x}_{\alpha}\vert^{2}} \\ \nonumber
\textrm{Where,} \quad
\boldsymbol{b}(\boldsymbol{x}) &=& \sum_{\alpha} \boldsymbol{b}(\boldsymbol{x}_{\alpha})\delta^{2}(\boldsymbol{x}-\boldsymbol{x}_{\alpha})
\end{eqnarray}

is the dislocation vector field. In the long wavelength limit, the r.m.s. fluctuations in $\tilde{\eta}$ behave as 

\begin{equation}\label{Defect2G}
\lim_{\boldsymbol{k}\rightarrow \boldsymbol{0}} \langle \vert \tilde{\eta}(\boldsymbol{k})\vert^{2}\rangle \propto 
\frac{\xi^{2}}{\vert \boldsymbol{k} \vert^{2} }.
\end{equation}

This is because, above the transition temperature, the unbinding dislocations interact via a screened Coulomb potential 
(See \ref{AppD} for details). Therefore, the thermal correlations in the Herringbone phase due to dislocations will read as

\begin{equation}\label{Defect2H}
\langle \Phi(\boldsymbol{x}) \Phi^{*}(\boldsymbol{0})\rangle \propto \vert \boldsymbol{x}\vert^{-n_{\Phi}}, \quad 
n_{\Phi} = \frac{2}{\pi c\xi^{2}_{\Phi}}.
\end{equation}

We observe that the correlations display algebraic decay (quasi-long-range order) rather than exponential decay. 
This indicates that in a two-dimensional system, or a quasi-two-dimensional state (such as confinement), the 
transition mediated by dislocations does not result in a completely disordered state. Instead, their unbinding 
leads to the formation of an orientationally ordered phase, known as the Herringbone phase or the Hexatic phase. 
The transition from either the Herringbone phase or the Hexatic phase to a completely disordered state can be 
aided by vortices, and this transition is of second order. However, this holds only if the system is strictly 
two-dimensional. In three dimensions, this transition is known to be first-order, as confirmed by experiments 
\cite{zammit1}. 

\section{Conclusion}\label{Sec6}

In the light of the Imry-Ma argument, we conclude that in dimension $d = 2$, the confined
$R_{I}$ phase will be unstable in the presence of quenched static random disorder.
Furthermore, we find that thermal fluctuations in the Herringbone order in the $X$-phase,
and the Hexatic order in the $R_{II}$-phase result in a quasi long range order, rather than
true long range order in the rotator phases. Lastly, we observe that the unbinding of dislocations 
results in the creation of an orientationally ordered Herringbone or Hexatic phases. 

\section{Acknowledgements}

We would like to express our gratitude to all the anonymous referees for their valuable suggestions. 
S.K.G. thanks the entire Ongil team for their constant technical and moral support.

\section{Appendix}\label{Sec7}

\subsection{The bulk and surface free energy of the $R_{I}-R_{II}$ phases, and the density of defects}\label{AppA}

For a $d$-dimensional system, the area $\mathcal{A} \propto L^{(d-1)}$. Let $s$ be the variable that 
measures the spatial variation of the domain wall separating $R_{I}$ and $R_{II}$. The direction of $s$ 
is perpendicular to $\mathcal{A}$, and the $d$-dimensional volume measure $d^{d}\boldsymbol{x} = \mathcal{A}ds$. 
To calculate the energy of this wall in $d$ dimensions, we will consider the gradient term, also 
known as the surface energy along with the confinement term in \eqref{Confinement1A}.

\begin{equation}\label{R12B}
H_{w}[\xi,\sigma] = \int_{0}^{L_{\Delta}} d^{d}\boldsymbol{x}
\bigg[\frac{c}{2} \vert \vec{\nabla}\xi(\boldsymbol{x}) \vert^{2} - \sigma(\boldsymbol{x})\xi(\boldsymbol{x})\bigg]
\end{equation}

If the value of the order parameter in the $R_{I}$ phase is $\xi_{0}$, then

\begin{equation}\label{R12C1}
\xi(\boldsymbol{x}) = \xi(x_{1},x_{2},..,x_{d-1},s) = \xi_{0}\Theta_{H}(s),    
\end{equation}

Where $\Theta_{H}(s) = 1, (s > 0)$, and zero otherwise, is the Heaviside or the step function defined in the regions 
where the domain boundaries exist. And its derivative is  

\begin{equation}\label{R12C2}
\vec{\nabla} \xi(\boldsymbol{x}) = \xi_{0}\frac{d\Theta_{H}(s)}{ds} = \xi_{0}\delta(s).
\end{equation}

a $\delta$-function defined in the small $\varepsilon$-neighbourhood of the domain boundary ($s \in [-\varepsilon/2,\varepsilon/2]$) 
(see FIG.\ref{Domain_Wall1}). The surface energy of the wall is then given by

\begin{eqnarray}\label{R12D}
\mathcal{E}_{w}(L_{\Delta}) &=& \frac{c \xi_{0}^{2}L_{\Delta}^{d-1}}{2a^{2}}\int_{s-\varepsilon/2}^{s+\varepsilon/2} ds \delta(s) \\ \nonumber
\textrm{or,} \quad
\mathcal{E}_{w}(L_{\Delta}) &=& \frac{c\xi^{2}_{0}}{2a^{2}} L_{\Delta}^{d-1}
\end{eqnarray}

Where $a$ is the lattice constant. Similarly, the confinement or the volume energy is given by

\begin{equation}\label{R12E}
\int_{0}^{L_{\Delta}} d^{d}\boldsymbol{x}\sigma(\boldsymbol{x})\xi(\boldsymbol{x}) = \xi_{0} \sigma L_{\Delta}^{d},
\end{equation}

and the corresponding r.m.s. fluctuations per domain wall is given by

\begin{equation}\label{R12G}
\mathcal{E}_{b}(L_{\Delta}) = -\xi_{0}\sqrt{\langle \sigma^{2} \rangle} L^{d/2}_{\Delta}.
\end{equation}

This is the bulk energy of the system and the total energy is given by

\begin{equation}\label{R12H}
\mathcal{E}(L_{\Delta}) = \frac{1}{2}c\xi^{2}_{0} L^{d_{c}/2}_{\Delta} - \xi_{0}\sqrt{\langle \sigma^{2} \rangle} L^{d/2}_{\Delta}.
\end{equation}

Where $d_{c} = 2(d-1)$. To find the density of the defects when the system is in its lowest energy state, we differentiate 
\eqref{R12H} w.r.t $L_{\Delta}$ and equate it to zero and find the following expression for the density of defects  

\begin{equation}\label{R12I}
\frac{L_{\Delta}}{a} = \bigg(\frac{c\xi_{0}d_{c}}{2d\sqrt{\langle \sigma^{2} \rangle}}\bigg)^{\frac{2}{2-d_{c}}}
\end{equation}

\subsection{Mean-field correlation length of domain walls}\label{AppB}

\begin{figure}[ht]
\centering
\subfigure[]{\includegraphics[width=0.3\textwidth]{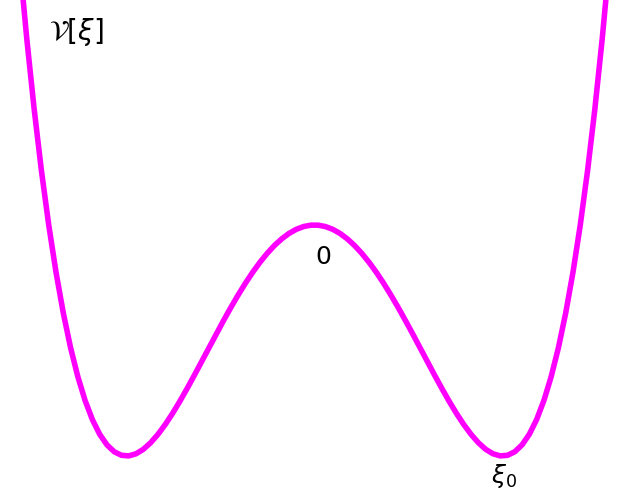}}\quad
\subfigure[]{\includegraphics[width=0.2\textwidth]{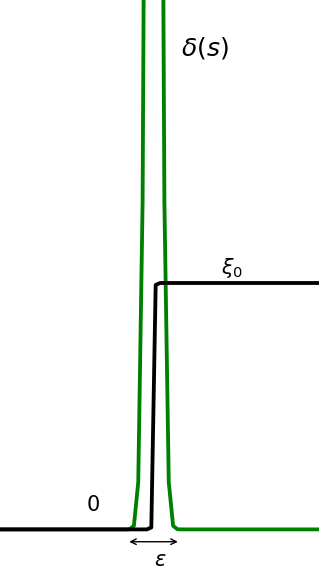}}
\caption{ (a) The double well potential where $\mathcal{V}[\xi = 0] = \alpha \xi^{2}_{0}/4$, and $\mathcal{V}[\xi = \xi_{0}] = 0$ ; 
(b) Plots for the order-parameter (black), and its derivative (green) for the domain wall.}
\label{Phi4}
\end{figure}

In the previous section, we assumed that the order parameter $\xi$ had a spatial profile resembling a step function. 
However, in general, the shape of the domain wall is determined by the system's free energy. This free energy dictates 
a characteristic length scale that is dependent on the parameters of the system. The typical size of the domain walls 
is set by the mean-field correlation length $\zeta_{d}$. In order to obtain an expression for $\zeta_{d}$, we ignore 
the quenched disorders ($\sigma = 0$), and consider only the $\xi^{2}$ and the $\xi^{4}$ terms ($\beta = 0)$ in the 
G-L functional in \eqref{Confinement1A}(FIG.\ref{Phi4}).  

\begin{equation}\label{AppendixA1a}
\frac{H[\xi]}{\mathcal{A}} = \int ds\bigg[\frac{c}{2}\bigg(\frac{d\xi(s)}{ds}\bigg)^{2} + \mathcal{V}[\xi(s)]\bigg]
\end{equation}

\begin{equation}\label{AppendixA1b}
\textrm{Where} \quad \mathcal{V}[\xi] = \frac{\alpha(\xi^{2} - \xi_{0}^{2})^{2}}{4\xi^{2}_{0}}, 
\quad \textrm{and} \quad \xi_{0} = \sqrt{\frac{\alpha}{\gamma}} \quad (\xi \geq 0). 
\end{equation}

The static Euler-Lagrange equation is obtained by calculating the functional derivative of \eqref{AppendixA1b} w.r.t. $\xi$ 

\begin{equation}\label{AppendixA1d}
\frac{\delta}{\delta \xi} \bigg(\frac{H[\xi]}{\mathcal{A}}\bigg) 
= -c\frac{d^{2}\xi(s)}{ds^{2}} + \frac{d\mathcal{V}[\xi(s)]}{ds} = 0
\end{equation}

Integration w.r.t $s \in [s_{0},s]$ gives us the following formal expression. 

\begin{equation}\label{AppendixA1e}
s-s_{0} = \int_{\mathcal{V}[\xi(s_{0})]}^{\mathcal{V}[\xi(s)]} \frac{d\xi}{\sqrt{2\mathcal{V}[\xi]/c}}
\end{equation}

Substituting the R.H.S. of \eqref{AppendixA1b}, and integrating w.r.t. $\xi$, we find the order-parameter (See FIG.\ref{Domain_Wall1}(b))

\begin{equation}\label{AppendixA1f}
\xi(s) = \xi_{0}\tanh\bigg(\frac{s-s_{0}}{\sqrt{2}\zeta_{d}}\bigg),  
\end{equation}

and the integrand in \eqref{AppendixA1a}, which is the energy density $\epsilon(s)$ (See FIG.\ref{Domain_Wall1}(a)) is given by

\begin{equation}\label{AppendixA1g}
\mathcal{E}(s) = \frac{\alpha\xi^{2}_{0}}{2}\sech^{4}\bigg(\frac{s-s_{0}}{\sqrt{2}\zeta_{d}}\bigg). 
\end{equation}

The quantity $\zeta_{d} \equiv \sqrt{c/\alpha}$, has the dimension of length, and is identified as the 
mean-field correlation length. 

\subsection{The bulk and surface free energy of Herringbone, $R_{II}$ phases, and the density of defects}\label{AppC}

We have seen in \ref{Sec3}, the order parameter for the Herringbone phase and the $R_{II}$ phase is a general complex 
scalar field $\Theta(\boldsymbol{x}) = \Theta_{0}e^{i\theta(\boldsymbol{x})}$ (N = 2). Following the arguments presented 
in \ref{AppA}, the surface energy density is given by       

\begin{eqnarray}\label{XYmodel0}
\frac{1}{2}c_{\Theta}\vert \vec{\nabla} \Theta(\boldsymbol{x})\vert^{2} &=& 
\lim_{\vert \boldsymbol a \vert \rightarrow 0} \quad 
\frac{1}{2}c_{\Theta}\Theta_{0}^{2} \bigg\vert \frac{e^{i\theta(\boldsymbol{x+a})} - e^{i\theta(\boldsymbol{x})}}{\boldsymbol{a}} \bigg\vert^{2}. 
\\ \nonumber
\textrm{Or,} \quad 
\frac{1}{2}c_{\Theta} \vert \vec{\nabla} \Theta(\boldsymbol{x})\vert^{2} &\approx& 
\frac{1}{2}c_{\Theta}\Theta_{0}^{2}\vert \vec{\nabla} \theta(\boldsymbol{x}) \vert^{2}.  
\end{eqnarray}

Where $c_{\Theta}$ is the stiffness constant of the rotators. Let $\theta(\boldsymbol{x})$ be the angle between the rotators and the external 
field $h(\boldsymbol{x})$. Then the bulk energy density will be proportional to the real part of $\Theta(\boldsymbol{x})$, and read as 

\begin{equation}\label{XYmodel1}
-\Theta_{0}h(\boldsymbol{x})\cos\theta(\boldsymbol{x}). 
\end{equation}

Combining \eqref{XYmodel0} and \eqref{XYmodel1}, and assuming that the spatial variation of the wall is along 
$s \in [0,L_{\Delta}]$, the free energy of the domain wall in this case is given by  

\begin{equation}\label{XYmodel2}
H[\theta] = 
\mathcal{A}\int_{0}^{L_{\Delta}} ds\bigg[ \frac{\rho_{\Theta}}{2} \bigg(\frac{d\theta(s)}{ds}\bigg)^{2} - \Theta_{0}h(s)\cos\theta(s) \bigg] 
\end{equation}

Where $\mathcal{A} \propto L_{\Delta}^{d-1}$, and $\rho_{\Theta} = c_{\Theta}a^{2}\Theta_{0}^{2}$, is the stiffness constant. 
To obtain an expression for the density of defects, let us assume that $\theta$ varies linearly with $s$, or $\theta = 2\pi s/L_{\Delta}$. 
Then the surface energy will read as 

\begin{equation}\label{XYmodel3}
\mathcal{E}_{s}(L_{\Delta}) = \frac{\rho_{\Theta}}{2}L_{\Delta}^{d-1} \int_{0}^{L_{\Delta}} ds \bigg(\frac{2\pi}{L_{\Delta}}\bigg)^{2} 
= 2\pi^{2}\rho_{\Theta}L_{\Delta}^{d-2}.  
\end{equation}

Secondly, if we replace $\cos\theta(s)$ by its thermal average $\langle \cos\theta(s) \rangle_{T} = \kappa(T)$, which is a 
temperature dependent dimensionless quantity, then the bulk energy per domain wall is given by

\begin{equation}\label{XYmodel4}
\mathcal{E}_{b}(L_{\Delta}) = -\kappa(T)\Theta_{0}\sqrt{\langle h^{2} \rangle} L^{d/2}_{\Delta}.
\end{equation}

Where $\sqrt{\langle h^{2} \rangle}$ is the r.m.s. fluctuations in $h$. The total energy, which is the sum of \eqref{XYmodel3} 
and \eqref{XYmodel4}, will then read as 

\begin{equation}\label{XYmodel5}
\mathcal{E}(L_{\Delta}) = 2\pi^{2}\rho_{\Theta}L^{d_{c}/2}_{\Delta} - \kappa(T)\Theta_{0}\sqrt{\langle h^{2} \rangle} L^{d/2}_{\Delta}.
\end{equation}

In the present case $d_{c} = 2(d-2)$, implying that the critical dimension is 4. Similar to \eqref{R12I}, the density of the 
defects in the system's lowest energy state is given by 

\begin{equation}\label{XYmodel6}
\frac{L_{\Delta}}{a} = \bigg(\frac{2\pi^{2}\rho_{\Theta}d_{c}}{\kappa(T)d\Theta_{0}\sqrt{\langle h^{2} \rangle}}\bigg)^{\frac{2}{2-d_{c}}}
\end{equation}

Similar to \eqref{AppendixA1b} in \ref{AppB}, the spatial profile of these domain walls can be obtained by calculating the 
functional derivative of \eqref{XYmodel2} w.r.t. $\theta$. The resulting static Euler-Lagrange equation is given by  

\begin{equation}\label{XYmodel7}
\frac{\delta}{\delta \theta} \bigg(\frac{H[\theta]}{\mathcal{A}}\bigg) =  
\frac{d^{2}\theta(s)}{ds^{2}} + \frac{\Theta_{0}h(s)}{\rho_{\Theta}}\sin\theta(s) = 0
\end{equation}

If $h(s) = h_{0}$, the amplitude value of the random field, then we can define a quantity $\lambda_{0} = \sqrt{\rho_{\Theta}/\Theta_{0}h_{0}}$, 
that has the dimensions of length. As a result, the system will exhibit domain walls characterized by a length that depends on the strength 
of the external random field. Equation \eqref{XYmodel7} resembles the equation of motion of a real pendulum, and its solutions will be 
described by Jacobi’s elliptic functions \cite{Abramowitz}.


\subsection{Elastic energy and thermal fluctuation calculations of crystalline solids}\label{AppD}

The strain tensor for crystalline solids in real and Fourier space in any dimensions is given by

\begin{eqnarray}\label{AppendixA1}
u_{ij}(\boldsymbol{x}) &=& \frac{1}{2}\bigg(\frac{\partial u_{i}}{\partial x_{j}} + \frac{\partial u_{j}}{\partial x_{i}} \bigg)   \\ \nonumber
u_{ij}(\boldsymbol{k}) &=& \frac{i}{2}( k_{j}u_{i} + k_{i}u_{j} )
\end{eqnarray}

If $\mu$ and $\lambda$ be the two elastic constants, then the elastic energy is given by

\begin{eqnarray}\label{AppendixA2}
\beta H &=& \frac{1}{2} \int d^{2}\boldsymbol{x} [2 \mu u_{ij}u_{ij} + \lambda u_{ii} u_{jj}]   \\ \nonumber
\beta H &=& \frac{1}{2} \int \frac{d^{2}\boldsymbol{k}}{(2\pi)^{2}}[2 \mu k_{i}k_{i}u_{j}u_{j} + (\mu + \lambda)(k_{i}u_{i})^{2}] \quad (\textrm{Fourier space})
\end{eqnarray}

The transverse and longitudinal energy for the phonons are given by

\begin{eqnarray}\label{AppendixA3}
\beta H_{T} &=& \frac{\mu}{2} \int \frac{d^{2}\boldsymbol{k}}{(2\pi)^{2}}k^{2}u^{2}_{T},   \\ \nonumber
\textrm{and} \quad
\beta H_{L} &=& \frac{2\mu+\lambda}{2} \int \frac{d^{2}\boldsymbol{k}}{(2\pi)^{2}} k^{2}u^{2}_{L}
\end{eqnarray}

respectively. According to the \textsl{equipartition theorem},

\begin{eqnarray}\label{AppendixA4}
&&\langle H_{T} \rangle = \langle H_{L} \rangle = \frac{1}{2}k_{B}T. \\ \nonumber
\textrm{Therefore,} \quad
&&\langle\vert u_{T}(\boldsymbol{k})\vert^{2}\rangle = \frac{1}{\mu k^{2}}, \quad \textrm{and} \quad
\langle\vert u_{L}(\boldsymbol{k})\vert^{2}\rangle = \frac{1}{(2\mu+\lambda) k^{2}}.
\end{eqnarray}
The displacement field correlation function will read

\begin{equation}\label{AppendixA5}
\langle u_{i}(\boldsymbol{k}) u_{j}(\boldsymbol{k'}) \rangle =
\frac{(2\pi)^{2}\delta(\boldsymbol{k}+\boldsymbol{k'})}{\mu \vert \boldsymbol{k}\vert^{2}}
\bigg[ \delta_{ij} - \frac{\mu+\lambda}{2\mu+\lambda}\frac{k_{i}k_{j}}{\vert \boldsymbol{k}\vert^{2}}\bigg].
\end{equation}

As a consequence of the above equation we will have the following relationship

\begin{eqnarray}\label{AppendixA6}
\langle [u_{i}(\boldsymbol{x}) - u_{i}(\boldsymbol{0})]^{2} \rangle
&=& \int \frac{d^{2}\boldsymbol{k}}{(2\pi)^{2}} \bigg(\frac{2-2\cos(\boldsymbol{k.x})}{\mu \vert \boldsymbol{k}\vert^{2}}\bigg)
\bigg[ \delta_{ii} - \frac{\mu+\lambda}{2\mu+\lambda}\frac{k_{i}k_{i}}{\vert \boldsymbol{k}\vert^{2}}\bigg], \\ \nonumber
&=& \frac{3\mu+\lambda}{\mu(2\mu+\lambda)}\frac{\ln(\vert \boldsymbol{x}\vert/a)}{\pi}.
\end{eqnarray}


\subsection{Density correlation function calculation of crystalline solids}\label{AppE}

The order parameter for the crystalline phase is the density function in the reciprocal space. 
Namely $\rho_{\boldsymbol{G}}(\boldsymbol{x})$, where $\boldsymbol{G}$ is any reciprocal
lattice vector. If $\boldsymbol{x}_{0}$ be a lattice vector then $\boldsymbol{G.x_{0}} = 2n\pi$, 
where $n$ is an integer. In two-dimensions or three-dimensions with stacked layers. 
The density-density correlation function obeys a power law given below.

\begin{eqnarray}\label{AppendixA7}
\langle \rho_{\boldsymbol{G}}(\boldsymbol{x}) \rho^{*}_{\boldsymbol{G}}(\boldsymbol{0})\rangle
&=& \langle \exp[i \boldsymbol{G}.\{\boldsymbol{u}(\boldsymbol{x}) - \boldsymbol{u}(\boldsymbol{0})\}]\rangle  \\ \nonumber
&=& \exp\bigg[-\frac{G_{i}G_{j}}{2}\langle \{u_{i}(\boldsymbol{x}) - u_{i}(\boldsymbol{0})\}\{u_{j}(\boldsymbol{x}) - u_{j}(\boldsymbol{0})\}\rangle\bigg]  \\ \nonumber
&=& \exp\bigg[-\frac{G_{i}G_{j}}{2} \int \frac{d^{2}\boldsymbol{k} d^{2}\boldsymbol{k'}}{(2\pi)^{2}}(e^{i\boldsymbol{k.x}}-1)(e^{i\boldsymbol{k'.x}}-1)
\langle u_{i}(\boldsymbol{k})u_{j}(\boldsymbol{k'})\rangle \bigg]
\\ \nonumber
&=& \exp\bigg[-\int \frac{d^{2}\boldsymbol{k}}{(2\pi)^{2}}\bigg(\frac{2-2\cos(\boldsymbol{k.x})}{\mu \vert \boldsymbol{k}\vert^{2}}\bigg)
\bigg( \frac{\vert \boldsymbol{G}\vert^{2}}{\mu} - \frac{\mu+\lambda}{2\mu+\lambda}\frac{(\boldsymbol{G.k})}{\vert \boldsymbol{k}\vert^{2}}\bigg)\bigg], \\ \nonumber
&\approx& \exp\bigg[-\frac{\vert \boldsymbol{G}\vert^{2}(3\mu+\lambda)}{2\mu(2\mu+\lambda)}\frac{\ln(\vert\boldsymbol{x}\vert/a)}{2\pi}\bigg]
= \bigg(\frac{a}{\vert \boldsymbol{x}\vert}\bigg)^{n_{\boldsymbol{G}}} \\ \nonumber
\textrm{Where,} \quad
n_{\boldsymbol{G}} &=& \frac{\vert\boldsymbol{G}\vert^{2}(3\mu+\lambda)}{4\pi\mu(2\mu+\lambda)}
\end{eqnarray}


\subsection{Correlation function calculation of the Herringbone order in the $X$-phase}\label{AppF}

\begin{eqnarray}\label{AppendixB1}
\langle \Phi(\boldsymbol{x}) \Phi^{*}(\boldsymbol{0})\rangle
&=& \phi^{2}\langle \exp(i2[\eta(\boldsymbol{x}) - \eta(\boldsymbol{0})])\rangle  \\ \nonumber
&=& \phi^{2}\exp\bigg[-\frac{2^{2}}{2}\frac{1}{4}
\langle [\boldsymbol{\nabla} \times \boldsymbol{u}(\boldsymbol{x}) - \boldsymbol{\nabla} \times \boldsymbol{u}(\boldsymbol{0})]^{2}\rangle\bigg]  \\ \nonumber
&=& \phi^{2}\exp\bigg[-\frac{1}{2} \int \frac{d^{2}\boldsymbol{k} d^{2}\boldsymbol{k'}}{(2\pi)^{2}}(e^{i\boldsymbol{k.x}}-1)(e^{i\boldsymbol{k'.x}}-1)
\underbrace{\epsilon_{aij}\epsilon_{akl}k_{i}k'_{k}}_{(\delta_{ik}\delta_{jl} - \delta_{il}\delta_{jk})k_{i}k'_{k}}
\langle u_{j}(\boldsymbol{k})u_{l}(\boldsymbol{k'})\rangle \bigg]
\\ \nonumber
&=& \phi^{2}\exp\bigg[- \frac{1}{2}\int \frac{d^{2}\boldsymbol{k}}{(2\pi)^{2}}(2-2\cos(\boldsymbol{k.x}))
\Big(\vert \boldsymbol{k}\vert^{2} \langle \vert \boldsymbol{u}(\boldsymbol{k})\vert^{2} \rangle -
\langle(\boldsymbol{k}.\boldsymbol{u}(\boldsymbol{k}))^{2}\rangle\Big)\bigg], \\ \nonumber
&=& \phi^{2}\exp\bigg[- \frac{1}{2\mu}\int \frac{d^{2}\boldsymbol{k}}{(2\pi)^{2}}\Big(2-2\cos(\boldsymbol{k.x})\Big)\bigg] \\ \nonumber
&\approx& \phi^{2}\exp\bigg(-\frac{1}{a^{2}\mu} \Big)
\end{eqnarray}

\subsection{Correlation function calculation of the dislocation vector fields and the corresponding Herringbone order in the $X$-phase}\label{AppG}

The singular part of the displacement vector fields can induce distortion in the Herringbone phase rotators, given by

\begin{equation}\label{AppendixC1}
\tilde{\eta}(\boldsymbol{x}) =
\frac{1}{2\pi} \int d^{2}\boldsymbol{x'} \frac{\boldsymbol{b}(\boldsymbol{x'}).(\boldsymbol{x}-\boldsymbol{x'})}{\vert\boldsymbol{x}-\boldsymbol{x'}\vert^{2}}.
\end{equation}

In Fourier space, 

\begin{equation}\label{AppendixC2}
\tilde{\eta}(\boldsymbol{k}) =
\frac{1}{2\pi} \int d^{2}\boldsymbol{x} \int d^{2}\boldsymbol{x'}e^{i \boldsymbol{k.x}}
\frac{\boldsymbol{b}(\boldsymbol{x'}).(\boldsymbol{x}-\boldsymbol{x'})}{\vert\boldsymbol{x}-\boldsymbol{x'}\vert^{2}}
= i\frac{\boldsymbol{k}.\boldsymbol{b}(\boldsymbol{k})}{\vert\boldsymbol{k}\vert^{2}}.
\end{equation}

The r.m.s. fluctuations in the phase is given by

\begin{eqnarray}\label{AppendixC3}
\langle \vert \tilde{\eta}(\boldsymbol{k})\vert^{2}\rangle &=&
\frac{k_{i}k_{j}}{\vert\boldsymbol{k}\vert^{4}}\langle b_{i}(\boldsymbol{k})b_{j}(\boldsymbol{k})\rangle.  \\ \nonumber
\textrm{Where}  \quad
\langle b_{i}(\boldsymbol{k})b_{j}(\boldsymbol{k})\rangle &=& \int d^{2}\boldsymbol{x} e^{i \boldsymbol{k.x}} \langle b_{i}(\boldsymbol{x})b_{j}(\boldsymbol{0})\rangle.
\end{eqnarray}

Above the transition temperature, if the unbinding dislocations interact via a screened Coulomb potential \eqref{Defect2E}, then

\begin{eqnarray}\label{AppendixC4}
\lim_{\vert \boldsymbol{x}\vert \rightarrow \infty} \langle b_{i}(\boldsymbol{x})b_{j}(\boldsymbol{0})\rangle &\propto&
\frac{\delta_{ij} \exp(-\vert \boldsymbol{x}\vert/\xi)}{\vert \boldsymbol{x}\vert}. \\ \nonumber
\textrm{Or,} \quad
\langle b_{i}(\boldsymbol{k})b_{j}(\boldsymbol{k})\rangle &\propto& \frac{\delta_{ij}}{ \vert \boldsymbol{k}\vert^2 + \xi^{-2}}. \\ \nonumber
\textrm{And,} \quad
\lim_{\vert \boldsymbol{k}\vert \rightarrow 0} \langle b_{i}(\boldsymbol{k})b_{j}(\boldsymbol{k})\rangle &\propto& \delta_{ij}\xi^{2}.
\end{eqnarray}

Substituting this in \eqref{AppendixC2}, we get \eqref{Defect2G}

\begin{equation}
\lim_{\boldsymbol{k}\rightarrow \boldsymbol{0}} \langle \vert \tilde{\eta}(\boldsymbol{k})\vert^{2}\rangle \propto 
\frac{\xi^{2}}{\vert \boldsymbol{k} \vert^{2} }.
\end{equation}

Using this result, the thermal correlations in the Herringbone phase due to the dislocation fields will read as

\begin{eqnarray}\label{AppendixC5}
\langle \Phi(\boldsymbol{x}) \Phi^{*}(\boldsymbol{0})\rangle
&\propto& \langle \exp(i2[\tilde{\eta}(\boldsymbol{x}) - \tilde{\eta}(\boldsymbol{0})])\rangle  \\ \nonumber
&=& \exp\bigg[-\frac{2^{2}}{2}\langle [ \tilde{\eta}(\boldsymbol{x}) - \tilde{\eta}(\boldsymbol{0})]^{2}\rangle\bigg]
= \vert \boldsymbol{x}\vert^{-n_{\Phi}}. \\ \nonumber
\textrm{Where,} \quad
n_{\Phi} &=& \frac{2}{\pi c\xi^{2}_{\Phi}}
\end{eqnarray}


\noindent
{\bf CRediT authorship contribution statement}

SKG: Calculations, developing theory, Writing - original draft, review and editing.
PKM: Calculations, developing theory, Writing - original draft, review and editing.

\noindent
{\bf Declaration of Competing Interest}

The authors declare that they have no known competing financial
interests or personal relationships that could have appeared to influence
the work reported in this manuscript.

\noindent
{\bf Data availability}

Data sharing is not applicable to this article as no new data were
created or analyzed in this study.

\noindent
{\bf Funding}

This research did not receive any specific grant from funding
agencies in the public, commercial, or not-for-profit sectors.

\end{document}